\documentclass[12pt,showkeys,pra,superscriptaddress,floatfix,nofootinbib,longbibliography]{revtex4-2}

\usepackage{graphicx}
\usepackage{color}
\usepackage{amsfonts}
\usepackage{eurosym}

\renewcommand{\d}{{\rm d}}

\newcommand {\E}  {\varepsilon}
\newcommand {\om} {\omega}
\newcommand {\Om} {\Omega}

\newcommand {\Ld}   {L_{\text{d}}}

\newcommand {\calE} {{\cal E}}

\begin{document}

\title{Intensive gamma-ray light sources
based on oriented single crystals}

\author{Gennady B. Sushko}
\affiliation{MBN Research Center, Altenh\"{o}ferallee 3, 60438 Frankfurt am Main, Germany}

\author{Andrei V. Korol}
\email[]{korol@mbnexplorer.com}
\affiliation{MBN Research Center, Altenh\"{o}ferallee 3, 60438 Frankfurt am Main, Germany}

\author{Andrey V. Solov'yov}
\email[]{solovyov@mbnresearch.com}
\affiliation{MBN Research Center, Altenh\"{o}ferallee 3, 60438 Frankfurt am Main, Germany}

\begin{abstract}
The feasibility of gamma-ray light sources based on
the channeling phenomenon of ultrarelativistic electrons
and positrons in oriented single crystals is
demonstrated using rigorous numerical modeling.
Case studies are presented for 10 GeV and sub-GeV
$e^{-}/e^{+}$ beams
incident on $10^{-1}-10^0$ mm thick diamond and silicon crystals.
It is shown that for moderate values of the beam average current
($\lesssim 10$ $\mu$A)
the average photon flux in the energy range $10^0-10^2$ MeV
emitted within the $10^1-10^3$ $\mu$rad cone
and 1 \% bandwidth can be on
the level of $10^{10}$ photon/s for electrons and
$10^{10}-10^{12}$ photon/s for positrons.
These values are higher than the fluxes available at modern
laser-Compton gamma ray light sources.
\end{abstract}

\maketitle

\section{Introduction  \label{Introduction}}

Powerful gamma-ray light sources (LSs)
operating in the MeV to GeV photon
energy range have a number of exciting potential applications in
a broad range of fields: fundamental science, industry, biology,
and medicine
\cite{AlbertThomas_PlasmaPhysContrFusion_v58_103001_2016,
Rehman_EtAl_ANE_v105_p150_2017,
Kraemer_EtAl-ScieRep_v8_p139_2018,
KorolSolovyov:EPJD_CLS_2020,
NextGenerationGammaRayLS2022,
CLS-book_2022}.
In this paper we propose a novel intensive gamma-ray LS
based on channeling radiation emitted by ultra-relativistic
electrons and positrons in thick single
crystals.
It is shown that photon fluxes generated in such systems can be
higher than achievable at modern
laser-Compton gamma-ray LS.

One of the existing methods for generating energetic
highly intensive
quasi-monoenergetic gamma rays with variable photon energies
relies on the laser-Compton scattering phenomenon.
In this process a low-energy laser photon experiences
back-scattering from an ultra-relativistic electron
acquiring the energy increase proportional to
the squared Lorentz factor
$\gamma = \E/mc^2$ \cite{Federici-EtAl:NuovoCim_vol59B_247_1980}.
This method has been used for producing gamma-rays in a broad,
$10^2$ keV  -- $10^0$ GeV, energy range.
Reviews on Compton gamma-ray beams and some of the commissioned
facilities are available
\cite{AlbertThomas_PlasmaPhysContrFusion_v58_103001_2016,
Krafft-Priebe:RevAccScieTechnol_vol3_p147_2010,
Sei-EtAl:ApplSci_vol10_p1418_2020,NextGenerationGammaRayLS2022,
Wu-EtAl:PRL_vol96_224801_2006}.
Among these facilities, to be mentioned is
the High Intensity Gamma-ray Source (HIGS)
\cite{Wu-EtAl:PRL_vol96_224801_2006} based on the Compton scattering
of the free-electron laser photons.
The HIGS facility generates a high-flux and nearly monochromatic gamma-rays
within the energy range from 1 to 100 MeV.

Intensive gamma-rays can be generated by means of
Crystal-based Light Sources (CLS) that can be set up by exposing
oriented crystals of different geometry (linear, bent,
periodically bent) to ultrarelativistic beams of electrons and
positrons \cite{KorolSolovyov:EPJD_CLS_2020,CLS-book_2022}.
Practical realization of CLSs is a subject of
the current European H2020
project N-LIGHT  \cite{N-Light} and Horizon  Europe
EIC-Pathfinder-2021
project TECHNO-CLS \cite{TECHNO-CLS}.
Construction of a CLS is a challenging task, which combines
development of technologies for crystal samples preparation,
extensive experimental programme that includes
design and manipulation of particle beams, detection and
characterization of the radiation,
as well as theoretical analysis and advanced  computational
modelling for the characterisation of CLSs.

The feasibility of CLSs based on the channeling phenomenon
of ultra-relativistic charged particles in periodically bent crystals
has been  demonstrated recently by means of numerical modeling
\cite{SushkoKorolSolovyov:EPJD_v76_166_2022,KorolSolovyov:NIMB_v537_p1_2023}.
In such systems, a projectile emits, in addition to the
channeling radiation (ChR) \cite{ChRad:Kumakhov1976}, the undulator-like
radiation as a result of the modulation of its trajectory due to the
periodic bending.
Case studies presented have shown that the
peak brilliance of radiation in the photon energy range
$\hbar\om =10^0$ MeV -- $10^0$ GeV
can be as high as achievable at modern synchrotron facilities
but at much lower photon energies.

The advantage of using periodically bent crystals is that
by varying the amplitude and period of bending one can optimize
the CLS characteristics
for given parameters of a beam.
This allows for tuning the parameters of the
radiation matching them to the needs of a particular application.
However, operational efficiency of such CLSs strongly depends
on the quality of periodical bending \cite{ChannelingBook2014}.
To achieve needed high quality a broad range of correlated
research and technological activities is needed.
One of the challenging technological task concerns manufacturing of periodically crystals \cite{TECHNO-CLS}.
An overview of relevant technologies can be found in Refs.
\cite{KorolSolovyov:EPJD_CLS_2020,CLS-book_2022}.

In this paper we demonstrate that
powerful gamma-ray CLS can be
realized based on the ChR emitted by energetic positrons and
electrons channeling in oriented single crystals of macroscopic
sizes ranging from hundreds of microns up to few millimeters.
Although the phenomena of channeling and ChR in single crystals
are well-documented (see, e.g., reviews
\cite{Andersen-EtAl_AnnRev_v33_p453_1983,Uggerhoj_RPM2005}),
there have been no systematical studies of using such crystals
as CLSs.
To the best of our knowledge the characterization of X-ray CLSs
based on the ChR in the 10-100 keV photon energy range
has been performed in Refs.
\cite{WagnerEtAl-NIMB_v266_p327_2008,Brau_EtAl-SynchrRadNews_v25_p20_2012}
for electron beams of moderate energies, $\E=10-40$ MeV.
The characteristic energy of ChR scales with the beam
energy as $\E^{2}$.
Therefore, to achieve emission of gamma-photons with
$\hbar\om \gtrsim 1$ MeV and beyond the energy of the incident beam
must be in the range of several GeV.
%


As a rule, modern accelerator facilities operate at a fixed value of
$\E$ (or, at several fixed values)
\cite{ParticleDataGroup2018,ShiltsevZimmermann:RPM_v93_015006_2019,
Yakimenko-EtAl:PR-AB_v22_101301_2019}.
Therefore, CLSs based on single crystals are less tunable as compared
to those based on periodically bent crystals.
However, linear oriented crystals of very high quality
are much more accessible.
This makes a perspective to create the CLSs by their means very attractive.
In the exemplary case studies presented in Section \ref{CaseStudies} below we
characterize the CLS based on positron ($e^{+}$) and electron ($e^{-}$) channeling in
diamond and silicon crystal.
The case studies refer to $e^{\pm}$ beams of essentially different
energies that are available at
(i) SLAC National Accelerator Laboratory
with $\E = 10$ GeV \cite{Yakimenko-EtAl:PR-AB_v22_101301_2019,%
FACETII_Technical_Design_Rep-2016,FACETII_Conceptual_Design_Rep-2015},
(ii) Mainz Microtron (MAMI) operating in the sub-GeV energy range
\cite{BackeEtAl:NIMB_266_p3835_2008,BackeEtAl:JINST_v13_C04022_2018,%
BackeEtAl_EPJD_v76_150_2022,MazzolariEtAl:arXiv_2404.08459}.
It is demonstrated that the photon fluxes generated in the CLS
in the photon energy range from 1 MeV up to hundreds of MeV
are higher than achievable at the existing laser Compton
gamma-ray LS \cite{NextGenerationGammaRayLS2022}.

For further reference and comparison, we present
Table \ref{Compton-LS.Table}, which lists the main parameters of
several major operational laser Compton gamma-ray LS
and new development projects.
The data are taken from table 10 in Ref. \cite{NextGenerationGammaRayLS2022}.

\begin{table*}[h]
\caption{
Parameters of several Compton gamma-ray LSs which are either
operational or being developed for operation,
Ref. \cite{NextGenerationGammaRayLS2022}.
Listed are:
electron beam energy $\E$,
average beam current $\langle I \rangle$,
gamma-photon energy $\hbar\om$,
$\gamma$-beam bandwidth (BW) $\Delta \om/\om$,
total average photon flux $\langle N_{\rm ph} \rangle_{\rm tot}$
and
on-target average flux $\langle N_{\rm ph} \rangle_{\rm tg}$.
The last line ("Status") indicates
operation status or projected operation date.
}
\begin{tabular}{p{2.6cm}p{1.9cm}p{2.4cm}p{2.3cm}p{2.1cm}p{2.05cm}p{2.cm}}
\hline
Facility  &  HIGS   &LEPS/LEPS2 &NewSUBARU &UVSOR-III &SLEGS    & ELI-NP  \\
Location  &   US    & Japan     &  Japan   &  Japan   & China   & Romania \\
\hline
$\E$ (GeV)&0.24-1.2 &  8        &  0.5-1.5 &  0.75    & 3.5     &0.23–0.74\\
$\langle I \rangle$ (mA)
          & 10-120  &  100      &  300     & 300      &100-300  &     1   \\
$\hbar\om$  (MeV)
         &  1–100  &1300–2900  & 1–40     & 1–5.4    & 0.4–20  &  1–19.5 \\
$\Delta \om/\om$ (\%)
          & 0.8–10  &  $<15$    &  10      &  2.9     &  $<5$   &  $<0.5$  \\
$\langle N_{\rm ph} \rangle_{\rm tot}$ (ph/s)
          &$10^6$-$3\!\times\!10^{10}$
                    &$10^6$-$10^7$&$10^7$-$4\!\times\!10^7$
                                          & $10^7$   &$10^6$-$10^8$
                                                                & $10^{11}$ \\
$\langle N_{\rm ph} \rangle_{\rm tg}$ (ph/s)
          & $10^3$-$3\!\times\!10^9$
                    &$10^6$-$10^7$&$10^5$-$3\!\times\!10^6$
                                           &$4\!\times\!10^5$
                                                     &$10^5$-$10^7$
                                                                & $\sim 10^8$ \\
Status    &Since 1996
                    & Since 1999& Since 2005
                                          &Since 2015&Test run                                                                 & Under
                                                                \\
          &
                    &           &
                                          &          & in 2023 \cite{SLEGS_ComptonLS_2022}
                                                                &       construction
                                                                \\
\hline
\end{tabular}
\label{Compton-LS.Table}
\end{table*}


\section{Results and discussion  \label{CaseStudies}}

Commonly, LSs operating in the short wavelength range
are characterized in terms of number of photons and brilliance
(see, e.g.,
\cite{AlbertThomas_PlasmaPhysContrFusion_v58_103001_2016,
NextGenerationGammaRayLS2022,%
Seddon-EtAl_RepProgPhys_v80_115901_720_2017,%
Bostedt-EtAl_RMP_v88_015007_2016,Couprie:JElSpectRelPhen_v196_p3_2014}).
Both quantities are proportional to spectral distribution
$\d E(\theta_0)/\d(\hbar\om)$
of energy radiated (per particle) in the solid
angle $\Delta\Om \approx \pi \theta_0^2$,
where the emission cone $\theta_0\ll 1$ is assumed to be small.
The number of photons $N_{\rm ph}$ and brilliance $B$ of radiation
emitted per unit time
within a bandwidth (BW) $\Delta\om / \om$
are calculated as follows
\begin{eqnarray}
N_{\rm ph}
=
{\d E(\theta_0) \over \d(\hbar\om)}
{\Delta \om \over \om}
{I \over e}
=
6.25\times 10^{12}
{\d E(\theta_0) \over \d(\hbar\om)}
{\Delta \om \over \om}
I \mbox{[$\mu$A]},
\label{eq.01}\\
B
=
{10^{-3} N_{\rm ph} \over
(\Delta\omega/\omega) (2\pi)^2\calE_x\calE_y}
=
1.58\times 10^{8}
{\d E (\theta_0)\over \d(\hbar\om)}
{ I \mbox{[$\mu$A]}\over  \calE_x\calE_y} .
\label{eq.02}
\end{eqnarray}
where $e$ is the elementary charge,
$I$ stands for the beam current.
The numerical factors on the right-hand sizes
corresponds to the current measured in micro amperes ($\mu$A).
The emission cone $\theta_0$ is measured with respect to the direction of the
incident beam (the $z$ axis).
The factors
$\calE_{x,y}=\sqrt{\sigma^2+\sigma_{x,y}^2}
\sqrt{\phi^2+\phi_{x,y}^2}$
stand for the total emittance of the photon source.
Here  $\sigma_{x,y}$ and $\phi_{x,y}$ are the root mean square sizes and divergence
of the beam in the transverse directions;
$\phi=\sqrt{\Delta\Om/2\pi}= \theta_0/\sqrt{2}$
and $\sigma=\lambda/4\pi\phi$ is the 'apparent' source size
taken in the diffraction limit \cite{Kim2009}
($\lambda$ is the radiation wavelength).
On the right-hand side of (\ref{eq.02})
$\sigma,\sigma_{x,y}$ are measured in mm,
and $\phi, \phi_{x,y}$ in mrad
so that $B$ is given in
$\hbox{photons/s/mrad}^{2}\hbox{/mm}^{2}/0.1\,\%\,\hbox{BW}$.

The number of photons and brilliance scale linearly with $I$.
Table \ref{Compton-LS.Table} (see also Table \ref{Table.SM-02}
in Appendix \ref{Beams2018}) shows that the average beam currents
achievable at modern facilities are in the range of
$10^2-10^3$ milliamperes.
These values are too high to be used in the channeling experiments
due to the limited durability of the crystalline targets,  even if they
are as resistant as diamond.
Therefore, the operation of the CLS has to be considered at much lower
beam currents.
In the case studies presented below in this section both quantities,
$N_{\rm ph}$ and $B$, are calculated for $I=1$ $\mu$A.
For other values of the current, the quantities are rescaled accordingly.
Note also is that the number of photons is calculated for a BW
equal to 1 \%.

To simulate the trajectories of ultra-relativistic charged particles
in a crystalline medium the method of Relativistic classical
Molecular Dynamics (Rel-MD) has been used, which is implemented in
the multi-purpose computer package \textsc{MBN Explorer}
\cite{MBNExplorer_2012,MBN_ChannelingPaper_2013,MBNExplorer_Book}.
The algorithms implemented allow for modelling passage of
particles over macroscopic distances with atomistic accuracy
accounting for the interaction of a projectile with all atoms
of the environment, thus
going beyond the continuous potential model \cite{Lindhard}
which is the basis for several software packages
\cite{Backe_EPJD_v76_143_2022,Sytov_PR_AB_v22_064601_2019,%
SytovTikhomirov_NIMB_v355_p383_2015,BagliGuidi_NIMB_v309_p124_2013}.
An overview of the results on channeling and radiation of charged
particles in crystals
simulated by means of
commercially available scientific software
\textsc{MBN Explorer} and a supplementary
multitask software toolkit \textsc{MBN Studio}
\cite{SushkoEtAl_2019-MBNStudio}
is given in
\cite{KorolSushkoSolovyov:EPJD_v75_p107_2021,CLS-book_2022}.

At the crystal entrance, the transverse coordinates and velocities of
an incident particle are generated using the normal
distributions with the standard deviations $\sigma_{x,y}$ and
$\phi_{x,y}$.
The positions of the lattice atoms are also generated randomly
accounting for the thermal vibrations corresponding to the room temperature.
Thus, each simulated trajectory corresponds to
unique initial conditions and crystalline environment so that
all trajectories are statistically independent.
For each trajectory
the spectral distribution of emitted radiation
has been calculated utilizing the  quasi-classical formalism
\cite{Baier} which accounts for the quantum recoil.
The spectral distribution of radiation emitted
by the beam particles
has been obtained by averaging the individual spectra.
This result has been used in Eqs. (\ref{eq.01})-(\ref{eq.02})
to calculate the number of photons and brilliance.

\subsection{Case study I: 10 GeV positrons and electrons  \label{CaseStudy_1}}

The simulations have been performed for $\E=10$ GeV
positron and electron beams
incident on oriented diamond and silicon single crystals
along the (110) planar direction.

The crystals thickness $L$ along the incident beam direction
($z$ direction)
has been varied from 1 to 6 mm for positrons and
fixed at $L=200$ $\mu$m for electrons.
%
These values of $L$ were chosen not to exceed the dechanneling lengths $L_{\rm d}$ of the
particles.
The latter quantity can be estimated within the framework of the diffusion model for
the dechanneling process
\cite{BeloshitskyEtAl:RadEff_v20_p90_1973,Baier,BiryukovChesnokovKotovBook}.
Using the explicit formulae for $L_{\rm d}$ given in Ref. \cite{ChannelingBook2014}
(see Sections 4.3.1 and 6.1.1) one calculates
$L_{\rm d}\approx5$ and 6 mm for 10 GeV positron channeling in diamond(110) and silicon(110),
respectively, and $L_{\rm d}\approx 200$ $\mu$m for the electron channeling in either of the crystals.
These values correlate with the results of Rel-MD simulations presented in Ref.
\cite{KorolSushkoSolovyov:EPJD_v75_p107_2021}.
Therefore, a large fraction of the incident particles pass through the whole crystall
moving in the channeling mode and, thus, contributing to the emission of the channeling radiation.
%

The parameters of the beam
(transverse sizes $\sigma_{x,y}= 17, 61$ $\mu$m (positrons),
$7, 16$ $\mu$m (electrons) and divergence
$\phi_{x,y}=10,\, 30$ $\mu$rad for both beams)
used in the simulations, correspond to the FACET-II beams
\cite{Yakimenko-EtAl:PR-AB_v22_101301_2019,FACETII_Conceptual_Design_Rep-2015,%
FACETII_Technical_Design_Rep-2016}.
In the simulations,
the $y$-axis was aligned with the $\langle 110\rangle$ axis.
This choice ensures that a sufficiently large fraction of the
beam particles is accepted in the channeling mode at the crystal
entrance since $\phi_y$ is ca two times smaller than
Lindhard's critical angle $\Theta_{\rm L}=(2U_0/\E)^{1/2}$.
Using $U_0\approx20$ eV for the interplanar potential depth in both
diamond and silicon (110) channels one calculates
$\Theta_{\rm L}\approx 63$ $\mu$rad.
The beam divergence along the $x$
direction  is much smaller than
the natural emission angle
$\gamma^{-1} \approx 50$ $\mu$rad.

\begin{figure*}[ht]
\centering
\includegraphics[scale=0.58]{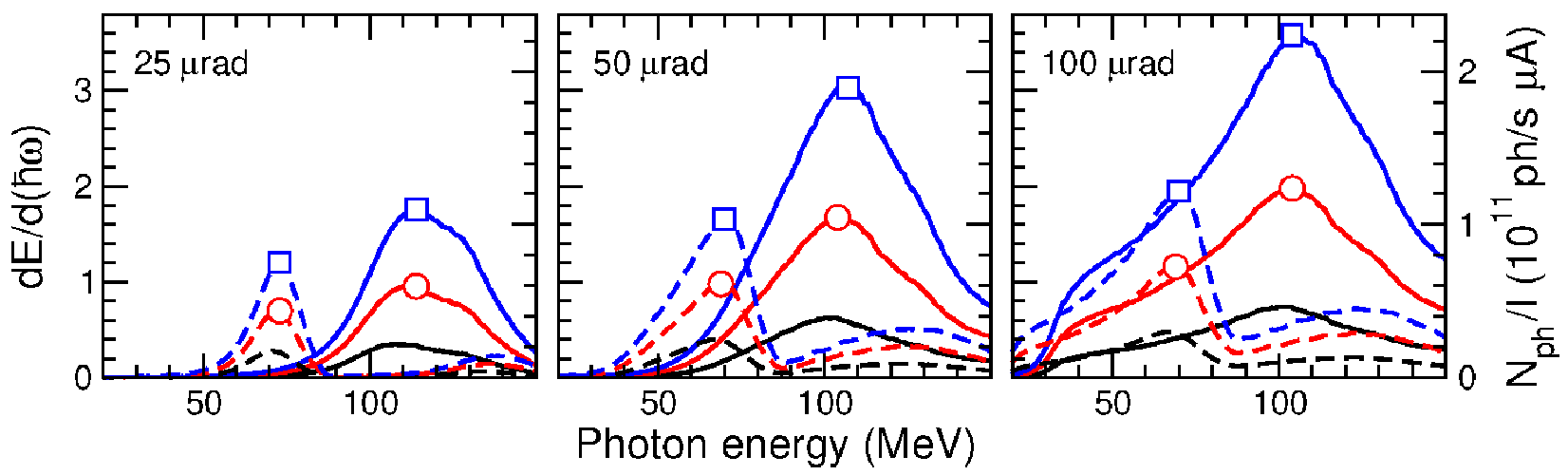}
\caption{
Spectral distribution of radiation (left axes)
and number of photons per second and per 1
$\mu$A of the beam current
(right axes) emitted by 10 GeV \textit{positrons}
incident along the (110) planar direction
in diamond (solid lines) and silicon (dashed lines) crystals.
The curves without symbols, with open circles and with open squares correspond to the crystal thickness
$L=1$, 3, and 6 mm, respectively.
Three graphs refer to different emission cones $\theta_0$ as
indicated.
The number of photons is calculated for the bandwidth
$\Delta \om /\om = 0.01$.
}
\label{Figure01.fig}
\end{figure*}

Figure \ref{Figure01.fig} presents the
results of calculations for the spectral distributions (left vertical
axes; absolute units) and for $N_{\rm ph}$ over $I$
(right vertical axes; units of $10^{11}$ photons/s/$\mu$A).
Three graphs refer to the emission cones
$\theta_0 =25$ $\mu$rad (left), 50 $\mu$rad (middle) and 100 $\mu$rad (right), which correspond approximately to
$1/2\gamma$, $1/\gamma$ and $2/\gamma$, respectively.
Solid curves show the dependencies calculated for the diamond crystals,
dashed curves -- to the silicon crystals.
In each graph and for each crystals the results shown correspond to
different values of the crystal thickness: $L=1$ mm (curves without
symbols), 3 mm (open circles), 6 mm (open squares).

Analysis of the  simulated trajectories
has shown that over 50 per cent of particles pass through
6 mm thick silicon and diamond crystals moving in the
channeling mode.
The spectral distributions shown in Fig. \ref{Figure01.fig}
are dominated by the peaks of ChR emitted by these
particles.
The peak intensities exceed the intensity of incoherent bremsstrahlung
emitted by the projectiles in the corresponding amorphous media by two
(for $\theta_0 =100$ $\mu$rad) to three
(for  $\theta_0 =25$ $\mu$rad) orders of magnitude,
see Figures \ref{Figure_SM-01.fig}-\ref{Figure_SM-03.fig} in Appendix \ref{Enhacement}.
For positrons, the channeling oscillations are nearly harmonic.
Therefore, to estimate the peak positions one can utilize the continuous
potential model \cite{Lindhard} assuming the harmonic approximation
for the inter-planar potential.
The corresponding estimates are presented in Appendix \ref{Estimates}.

The profile of the number of photons as a function of photon energy
is the same as for the spectral distribution since both quantities
differ by a numerical factor that includes the product of BW and
beam current.
The tick labels of the right vertical axes in Fig.
\ref{Figure01.fig} indicate the values of
$N_{\rm ph}$ scaled by the beam current measured in microamperes.
All data refer to BW $\Delta \om /\om =0.01$.

Nominal parameters of the 10 GeV positron FACET-II beam include
bunch charge $Q=1$ nC, $I_{\rm peak}=6$ kA and pulse repetition rate
$f=5$ Hz \cite{Yakimenko-EtAl:PR-AB_v22_101301_2019}.
Multiplying the right tick labels in
Fig. \ref{Figure01.fig}
by $I_{\rm peak}$ one obtains the \textit{peak values} of the number of
photons as high as $\sim 10^{21}$ ph/s.
%
In this case we note that it is known (at least for the diamond targets)
that ``\dots a diamond crystal bears no visible influence
from being irradiated by the final focus test beam\dots '' at the SLAC facility (see
Ref. \cite{Uggerhoj_RPM2005}, p. 1160).
It is therefore feasible to use these crystals as the sources of intense gamma rays.

Another essential quantity commonly considered for the gamma-ray LS
operating in the energy range beyond 1 MeV is the \textit{average}
number of gamma-photons,
$\langle N_{\rm ph} \rangle$ \cite{NextGenerationGammaRayLS2022}.
To calculate this quantity for the FACET-II beam
one multiplies $N_{\rm ph} / I$ by
the beam's average current $\langle I \rangle = fQ = 5\times10^{-3}$ $\mu$A.
Then, in the vicinity of the maxima ($\hbar\om_{\max} \approx 70$ MeV for silicon
and $\hbar\om_{\max} \approx 100-110$ MeV for diamond) one finds that
$\langle N_{\rm ph} \rangle$ from $10^8$ ph/s for both crystals
of thickness 1 mm and small emission cone (left graph in the figure)
up to $(0.5-1)\times10^9$ ph/s for $L=6$ mm and $\theta_{0}=100$ $\mu$rad
(right graph).
These values are comparable with the on-target photon flux
generated in this photon energy range by the HIGS facility
by using 0.24-1.2 GeV \textit{electron} beam of the average current
$10-120$ mA (see Table \ref{Compton-LS.Table}),
which is six to seven orders
of magnitude larger then $\langle I \rangle$ of the FACET-II
beam.
We note that several GeV \textit{positron} beams
of the average current $10^2-10^3$ mA are available at present,
see Table \ref{Table.SM-02} in
Appendix \ref{Beams2018}.
Therefore, even if considering
smaller currents
$\langle I \rangle\lesssim 10$ $\mu$A \cite{ChaikovskaEtAl:JINST_v17_P05015_2022}
it is feasible by means of oriented crystals
to achieve photon average fluxes
$\langle N_{\rm ph} \rangle \sim 10^{12}$
ph/s emitted within
narrow, $\theta_0=10^1-10^2$ $\mu$rad, cones.
These values of $\langle N_{\rm ph} \rangle$ are higher
then those expected in the
the next-generation gamma-ray source HIGS2 \cite{Wu_talk2020}.

\begin{figure*}
\centering
\includegraphics[scale=1.15]{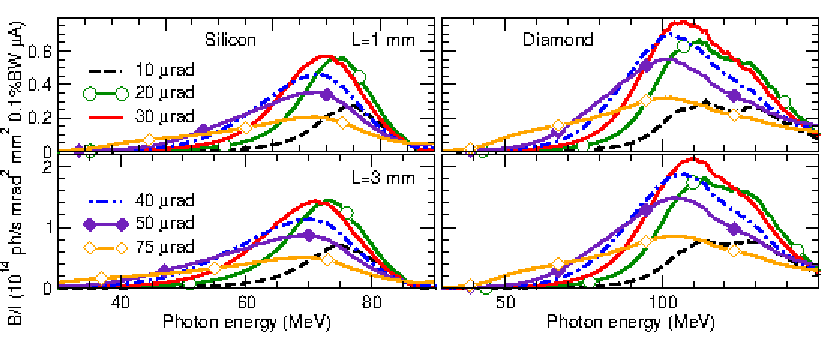}
\caption{
Brilliance of radiation per 1 $\mu$A
of the positron beam current
calculated for different emission cones $\theta_0$ as
indicated in common legends.
Left and right column correspond, respectively, to the
diamond and silicon crystals.
Upper row show the results for $L=1$ mm thick crystals,
lower row -- for $L=3$ mm.
All data refer to the bandwidth
$\Delta \om /\om = 0.01$.
}
\label{Figure02.fig}
\end{figure*}


Figure \ref{Figure02.fig} shows the brilliance of radiation (scaled by
the beam current measured in $\mu$A)
emitted in $L=1$ mm (top) and 3 mm (bottom) thick
silicon (left) and diamond (right) crystals.
In each graph the curves correspond to different emission cones
$\theta_0$ indicated in the common legends.
In contrast to spectral distribution, which is an
increasing function of the emission cone for all values of $\om$,
the brilliance exhibits a non-monotonous behaviour
due to the presence of the terms $\theta_0^2$ in the factors
$\calE_{x,y}$, see Eq. (\ref{eq.02}).
For both crystals the maximum peak values are achieved at
$\theta_0 \approx 30$ $\mu$rad, which is about two times less than
the natural emission $1/\gamma$.
Multiplying the curves by the peak current $I_{\rm peak}=6$ kA one
calculates the peak brilliance $B_{\rm peak}$, which characterizes the
photon emission by a single bunch of the beam:
$B_{\rm peak} \approx (3-8)\times 10^{23}$ ph/s/mrad$^2$mm$^2$/0.1 \% BW
at $\hbar\om \approx 70$ MeV (silicon)
and
$(4-10)\times 10^{23}$ ph/s/mrad$^2$mm$^2$/0.1 \% BW at
$\hbar\om \approx 100$ MeV (diamond).
These values of $B_{\rm peak}$ are comparable or higher that peak
brilliance of radiation emitted at the modern synchrotron facilities but
in much for lower, $10^1-10^2$ keV, photon energies
\cite{DuerrEtAl-IEEETransMagn_v45_p15_2009}.


\begin{figure*}[ht!]
\centering
\includegraphics[scale=0.285]{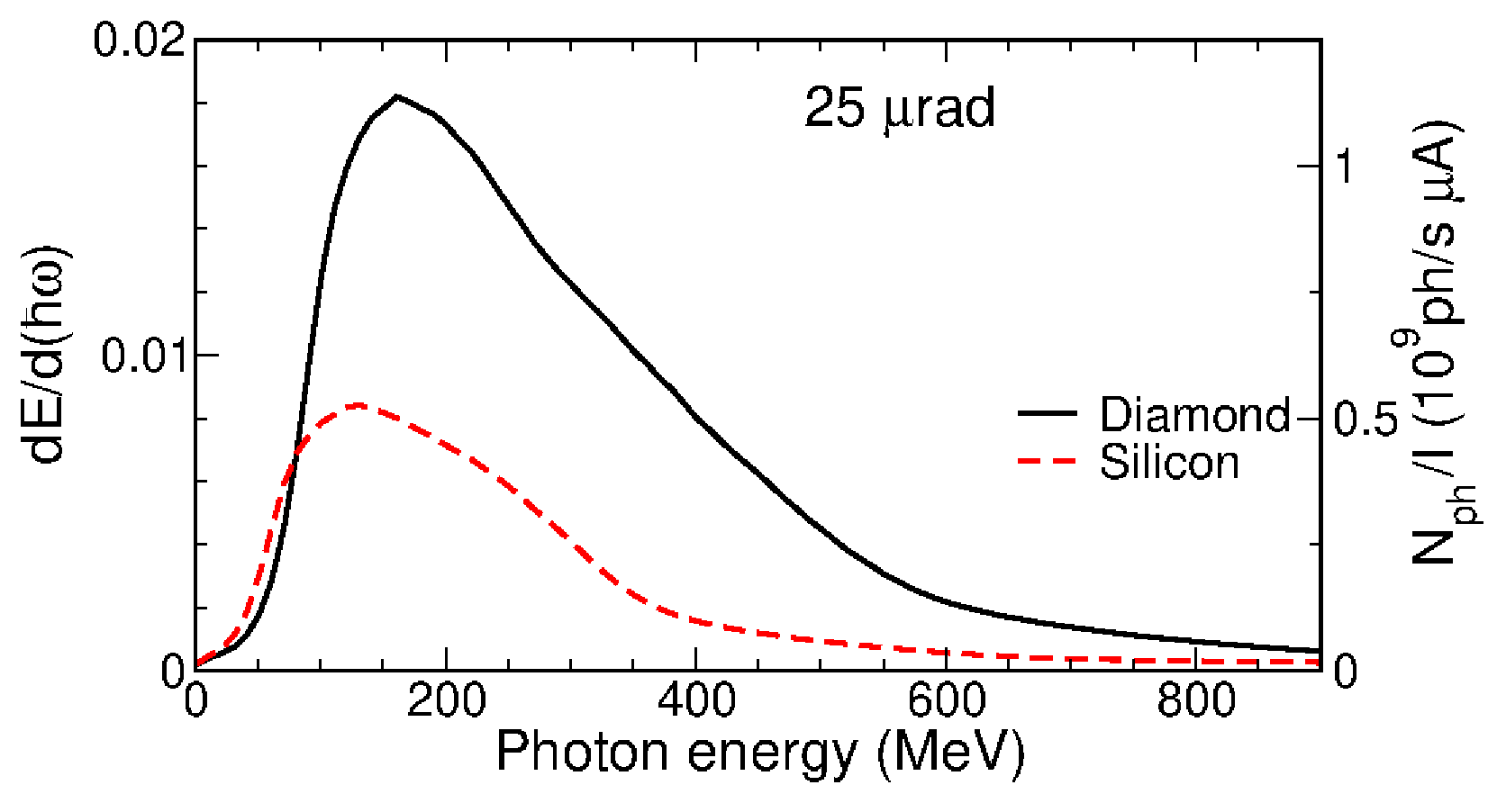}
\hspace{0.1cm}
\includegraphics[scale=0.285]{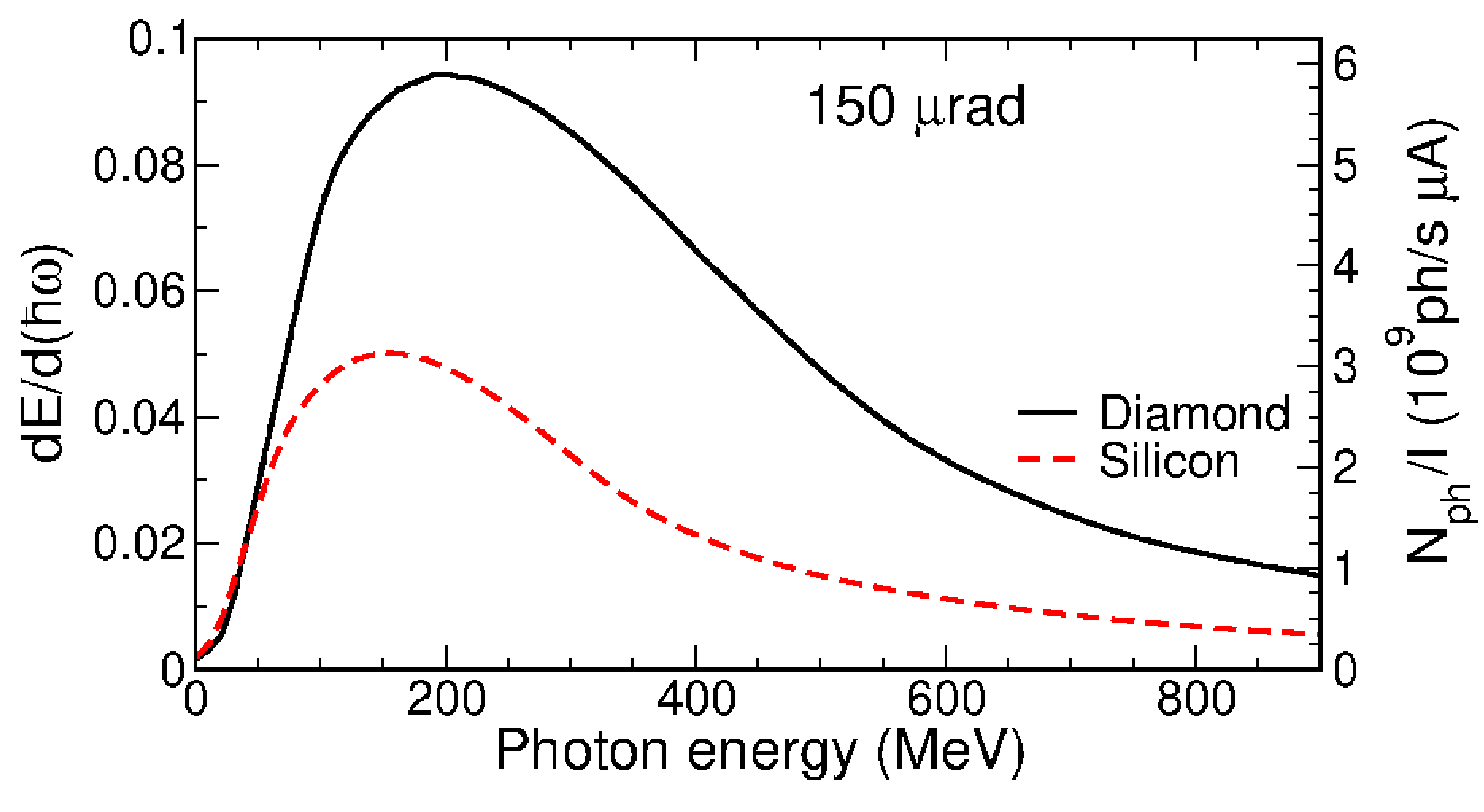}
\caption{Spectral distribution radiation
(left vertical axes)
and number of photons per second and per unit
current in $\mu$A
(right vertical axes) calculated for 10 GeV
\textit{electrons}
passing through $L=200$ $\mu$m thick diamond and silicon crystals
along the (110) planar direction.
Left and right graphs correspond to the emission cones
$\theta_0=25$ and 100 $\mu$rad, respectively.
The number of photons refers to the bandwidth
$\Delta \om /\om = 0.01$.
}
\label{Figure03.fig}%
\end{figure*}

The dechanneling phenomenon is the main reason that does
not allow for unrestricted increase in the intensity of ChR
by increasing the thickness $L$ of a crystalline target.
In the case of a planar channeling, the dechanneling length $\Ld$ of positrons
exceeds that of electrons of the same energy by more than an order of
magnitude (see, e.g., Ref. \cite{ChannelingBook2014}).
As a result, highly intensive ChR can be emitted by positrons
during their passage through much thicker crystals.
Additionally, channeling oscillations of positrons, being nearly
harmonic, lead to the undulator-like pattern of the ChR
spectral-angular distribution.
In particular, for each value of emission angle $\theta$ the
distribution consists of a set of narrow peaks (harmonics)
the energies of which are integer multiples of the fundamental
harmonic (see estimates in Appendix \ref{Estimates}).
In contrast, strong anharmonicity of the electron channeling
oscillations
leads to the distribution of the ChR over much broader energy interval,
see, e.g., Ref. \cite{Bak-EtAl:NuclPhysB_v251_p254_1985}.

However, taking into account that high-energy electron
beams are available worldwide at much larger number of facilities,
it is meaningful to characterize the CLS based on the electron channeling
in thinner crystals.

The simulations have been performed for the electron beam
incident along the (110) planar direction on 200 $\mu$m thick
diamond and silicon crystals.
This chosen value exceeds the dechanneling lengths 91 and
73 $\mu$m
of 10 GeV electrons
in  C(110) and Si(110) planar channels, respectively
Ref. \cite{KorolSolovyov:NIMB_v537_p1_2023}.
Spectral distributions of the radiated energy and of
the number of photons scaled by the current are shown in
Fig. \ref{Figure03.fig}, left and right panels of which
refer to the emission cones $25$ and 100 $\mu$rad.
Note that the values of $\d E/\d(\hbar\om)$
are multiplied by $10^3$.
The right vertical axes show the values
$N_{\rm ph}/I$ in units of $10^{9}$ ph/s/$\mu$A).
To calculate the average current of the FACET-II electron beam one
can use the data on the bunch charge 2 nC and repetition rate 30 Hz
 \cite{Yakimenko-EtAl:PR-AB_v22_101301_2019}:
$\langle I \rangle =6\times 10^{-2}$ $\mu$A.
Then, the expected values of the average photon flux emitted
within the cones $\theta_0=25-100$ $\mu$rad and
within the 1\% BW at $\hbar\om\lesssim 200$ MeV  are
$\langle N_{\rm ph} \rangle = (0.6-4)\times 10^{8}$ ph/s
for the diamond crystal and approximately two times less for the
silicon target.
The values of
$\langle N_{\rm ph} \rangle$ can be increased by orders
magnitudes considering the electron beam currents
$\langle I \rangle \sim 10^0-10^1$ $\mu$A
achievable at modern facilities
\cite{ShiltsevZimmermann:RPM_v93_015006_2019,NextGenerationGammaRayLS2022}.

\subsection{Case study II: Sub-GeV electrons and positrons \label{CaseStudy_2}}

Over the last two decades a number of channeling experiments with 195-855 MeV electron beams
have been carried out at the Mainz Microtron MAMI facility with various crystals (diamond, silicon, germanium) of different
geometry (linear, bent, periodically bent crystals) \cite{BackeEtAl:NIMB_266_p3835_2008,
Backe_etal:JPConfSer_v438_012017_2013,Backe_EtAl_2013,
Wistisen_etal_PRL_2014,
BackeLauth:NIMB_v355_p24_2015,
Bandiera_EtAl:PRL_v115_025504_2015,Mazzolari-EtAl_EPJC_78_p720_2018,
Haurylavets_EtAl-EPJPlus_v137_34_2022}.
A more recent advance concerns a design study that was performed for a
positron beam with an energy of 530 MeV to be realized at MAMI
\cite{BackeEtAl_EPJD_v76_150_2022}.
In the year 2024 the positron beamline has been completed and the
results of the first channeling experiments carried out with bent
silicon crystal were reported \cite{MazzolariEtAl:arXiv_2404.08459}.

In this Section we report on the results of computation of the CLS characteristics based on
the Rel-MD simulations of the channeling process and photon emission by 855 MeV electrons
and 500 MeV positrons incident on single diamond crystals.

For electrons two crystal orientations, --  along the (110) and (111)
planar directions, have been considered.
In the diamond cubic lattice the (111) interplanar spacing is
$2\sqrt{2}/3\approx 1.6$ times larger than that for the (110) planes.
The dechanneling length in the (111) channels
$L_{\rm d}\approx 19$ $\mu$m is larger by approximately the same factor
the value $L_{\rm d}\approx 12$ $\mu$m in the (110) channels
\cite{KorolSushkoSolovyov:EPJD_v75_p107_2021}.
Hence, one would expect more intensive ChR radiation in the case of
the (111) planar channeling.
The simulations were performed for the crystal thicknesses within the
range $L=50-200$ $\mu$m.
The beam divergence $\phi_{y}=10$ $\mu$rad
\cite{BackeEtAl:NIMB_266_p3835_2008} used in the
simulations is much smaller that Lindhard's critical angles
$\Theta_{\rm L} \approx 240$ and 225 $\mu$rad in the (111) and (110)
channels, respectively.

\begin{figure*}[h]
\centering
\includegraphics[scale=0.58]{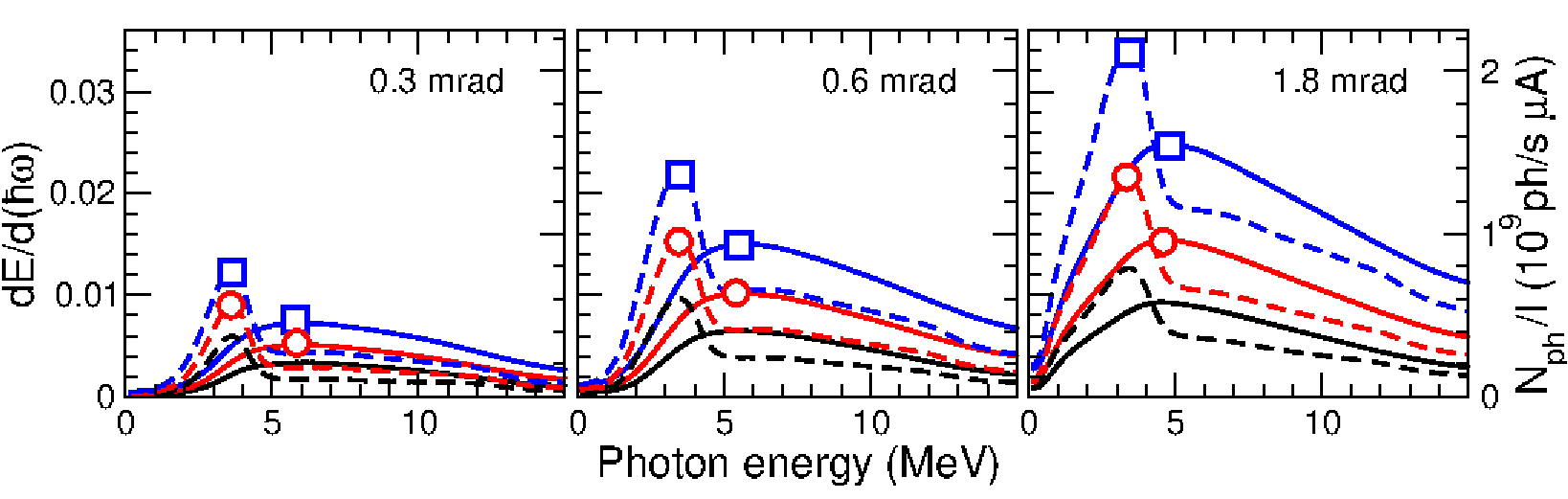}
\caption{
Spectral distribution of radiation (left axes)
and number of photons per second and per 1 $\mu$A of the beam current
(right axes) emitted by 855 MeV \textit{electrons}
incident on a diamond crystal along the (110) (solid curves) and
(111) (dashed curves) planar directions.
The curves without symbols, with open circles and with open squares correspond to the crystal thickness $L=50$, 100, and 200 $\mu$m, respectively.
Three graphs refer to different emission cones $\theta_0$ as
indicated.
The number of photons is calculated for the bandwidth
$\Delta \om /\om = 0.01$.
}
\label{Figure_e855-C-allL.fig}
\end{figure*}

Figure \ref{Figure_e855-C-allL.fig} shows the spectral distributions (left vertical
axes) and the number of photons per second scaled by the beam current
(right vertical axes; units of $10^{9}$ photons/s/$\mu$A).
Three graphs refer to the emission cones
$\theta_0 =0.3$, 0.6 and 1.8 mrad, which are equal approximately to
$1/2\gamma$, $1/\gamma$ and $3/\gamma$, respectively.
Solid curves present the dependencies calculated for the (110) planar orientation
of the crystal, dashed curves -- for the (111) orientation.
In each graph and for each crystals the results shown correspond to
different values of the crystal thickness: $L=50$ (curves without
symbols), 100 (circles), 200 $\mu$m (squares).
To be noted is that in the case of the (111) channeling the maxima
are narrower and more powerful, and are located at lower photon energies (about
3 MeV) as compared to $\hbar\om_{\max} \approx 5$ MeV for the (110) planar channeling.
Another feature clearly seen in all graphs is that the photon yield is not saturated
with the crystal thickness (i.e. does not become independent on $L$) despite the strong
inequality $L\gg L_{\rm d}$.
This is due to a high rate of the rechanneling events (i.e. capture of the overbarrier
particles into the channeling mode) that happen to the electrons inside the crystal \cite{KorolSushkoSolovyov:EPJD_v75_p107_2021}.

For the main goal of this paper it is important to stress high values of the photon flux
that can be generated.
Figure \ref{Figure_e855-C-allL.fig} shows that for the beam current of 1 $\mu$A
the number of photons within the energy range $2-15$ MeV is on the
level $10^{9}$ ph/s.
Increasing the average current to 10 $\mu$A (which is achievable at the MAMI facility
\cite{MAMI}) the number of photons can be as high as 10$^{10}$ ph/s,
i.e. exceeding the values achievable at most of the currently operating
Compton gamma-ray LS, see Table \ref{Compton-LS.Table}.

As part of the ongoing TECHNO-CLS project
\cite{TECHNO-CLS}, a new  positron beamline has been built at the
MAMI facility.
The beam energy is 530 $\pm$ 10 MeV and the divergence
$\sigma_{y}\approx 60$ $\mu$rad.
The intensity of the beam is currently low (a few thousand
positrons per second \cite{BackeLauth_Private_2024}),
but, nevertheless, it can already be used for channeling experiments
with crystals \cite{MazzolariEtAl:arXiv_2404.08459}.
The low intensity beam can also be used for the
experimental characterisation of the CLS prototype based on the
sub-GeV positrons.

\begin{figure*}[h]
\centering
\includegraphics[scale=0.58]{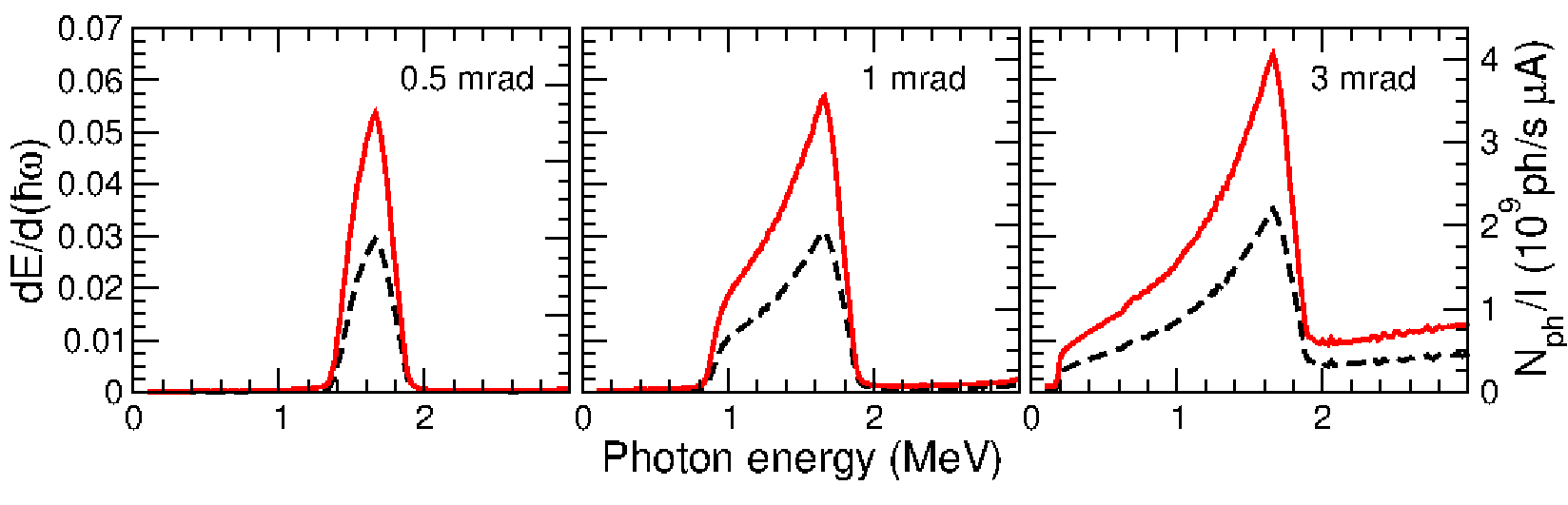}
\caption{
Spectral distribution of radiation (left axes)
and number of photons per second and per 1 $\mu$A of the beam current
(right axes) emitted by 530 MeV \textit{positrons}
incident on a diamond crystal along the (110)  planar direction.
Solid and dashed curves correspond to the crystal thickness 400 and 200 $\mu$m, respectively.
Three graphs refer to different emission cones $\theta_0$ as
indicated.
The number of photons is calculated for the bandwidth
$\Delta \om /\om = 0.01$.
}
\label{Figure_p500-C110-allL.fig}
\end{figure*}

The quoted values of $\E$ and $\sigma_{y}$ have been used to
in the simulations performed for the theoretical characterization of
the CLS based on positron channeling in
$L=200$ and 400 $\mu$m thick oriented single diamond(110) crystals.
Figure \ref{Figure_p500-C110-allL.fig} shows the results obtained
for the spectral distribution of the emitted radiation
and for the number of photons.
Three plots refer to the emission cones
$\theta_0 =0.5$, 1.0 and 3.0 mrad, which correspond
approximately to $1/2\gamma$, $1/\gamma$ and $3/\gamma$, respectively.
The dashed and solid lines show the dependencies calculated for
the smallest and largest values of $L$.
As expected, the peaks of the channeling radiation by positrons are
higher and much narrower than for electrons.
The radiation is emitted mainly in the photon energy range 1-2 MeV
(the lower bandwidth decreases as the emission cone increases).
For the average current of 1 $\mu$m the photon flux can be expected
on the level of several 10$^{9}$ ph/s.

\section{Conclusions  \label{Conclusions}}

In this paper, we have proposed and characterized novel intensive
gamma-ray LS based on the channeling radiation emitted
by ultra-relativistic electrons and positrons in oriented single
crystals.
The feasibility of such LSs is demonstrated by means of rigorous
numerical modeling.
The results of numerical simulations presented refer to
two specific case studies of the radiation emission (i)
by  10 GeV projectiles channeling in diamond and silicon crystals, and
(ii) sub-GeV electrons and positrons in the diamond crystal.
It is shown that for average beam current of the order of
10 $\mu$A the average flux of photons emitted within the
cone $\theta_0 \leq (2-3)/\gamma$ and
within 1 \% bandwidth can be as high as
$10^{12}$ ph/s within the energy range $\hbar\om =50-150$ MeV
for the 10 GeV positron beam and
$10^{10}$ ph/s within $\hbar\om =100-800$ MeV for the
10 GeV electron beam.
For the sub-GeV projectiles the fluxes of $10^{10}$ ph/s can be achieved at much
shorter crystals within the photon energy range
$\hbar\om \lesssim 10$ MeV.
%

The case studies presented clearly show that wih the help of CLS
it is possible to generate photon fluxes in the energy range
$10^0-10^2$ MeV higher than those
achievable with current (or planned to be commissioned in the future)
Compton gamma-ray sources, which use beam currents that are orders of
magnitude larger \cite{Wu_talk2020}.

For given parameters of an available beam (modality, energy,
emittance),
the photon flux and/or the brilliance of the CLS can be maximized by
varying the crystalline medium, the crystal thickness
and its orientation with respect to the beam as well as
utilizing different channeling regimes (planar or axial channeling).
This allows an optimal selection of the CLS to be used in a specific
experimental environment and/or tune the parameters of the emitted
to the needs of a particular application.

This paper focuses on the CLS based on linear crystals.
The exposure of oriented crystals of other geometries (bent
crystals, periodically bent crystals) opens up further
possibilities for the construction of
novel intensive gamma-ray CLS that can operate in a much wider
photon energy range, from hundreds of keV up to several GeV
\cite{CLS-book_2022}.
Construction of crystal-based LSs is a challenging task, which
combines development of technologies for crystal samples
preparation, extensive experimental programme that includes design
and manipulation of particle beams, detection and characterization
of the radiation, as well as theoretical analysis and advanced
computational modelling for the characterisation of CLSs.
The consortium of the ongoing TECHNO-CLS project
\cite{TECHNO-CLS} has all the necessary expertise to
carry out the outlined programme and to demonstrate the prototypes of the novel gamma-ray LSs.

\vspace*{0.2cm}

The work was supported by the European Commission through the N-LIGHT
Project within the H2020-MSCA-RISE-2019 call (GA 872196)
and the Horizon  Europe EIC-Pathfinder-Project TECHNO-CLS
(Project No. 101046458).
We acknowledge the Frankfurt Center for Scientific Computing (CSC) for
providing computer facilities.

\appendix

\section{Lepton beams parameters
\label{Beams2018}}

The data on positron and electron beams energy $\E$ and
average beam current $I$ are summarized in
Table \ref{Table.SM-02}.
The table compiles the data for the following facilities:
VEPP4M (Russia), BEPCII (China) \cite{ParticleDataGroup2018},
SuperKEKB (Japan)
\cite{2018_SuperKEKB},
SuperB (Italy) \cite{ParticleDataGroup2010}
(its construction was canceled \cite{SuperB-Cancelled}),
and DESYII test beam \cite{2019_DESYII_TestBeam}.

\begin{table*}[h]
\caption{
Parameters of positron ('p') and electron ('e') beams:
beam energy, $\E$,
average beam current $\langle I \rangle$.
In the cells with no explicit reference to either 'e' or 'p' the data refer to
both modalities.
}
\begin{tabular}{p{2.cm}p{2.cm}p{2.cm}p{2.3cm}p{2.cm}p{2.cm}}
\hline
 Facility  &VEPP4M&BEPCII & SuperKEKB       & SuperB   & DESY II \\
 Ref.      &\cite{ParticleDataGroup2018}
                                  & \cite{ParticleDataGroup2018}
                                             & \cite{2018_SuperKEKB} &
                                \cite{ParticleDataGroup2010,SuperB-Cancelled}
                                &\cite{2019_DESYII_TestBeam}\\
\hline
$\E$ (GeV)               &6    &1.9-2.3&{\rm p}: 4     &{\rm p}: 6.7  & 1-6 \\
                         &     &       &{\rm e}: 7     &{\rm e}: 4.2  &      \\
$\langle I \rangle$ (mA) &80   & 851   &{\rm p}: 3600  &{\rm p}: 2400 &\\
                         &     &       &{\rm e}: 2600  &{\rm e}: 1900 &\\
\hline
\end{tabular}
\label{Table.SM-02}
\end{table*}

\section{Enhancement of radiation in oriented crystals
\label{Enhacement}}

\begin{figure*} [h]
\centering
\includegraphics[width=16.4cm]{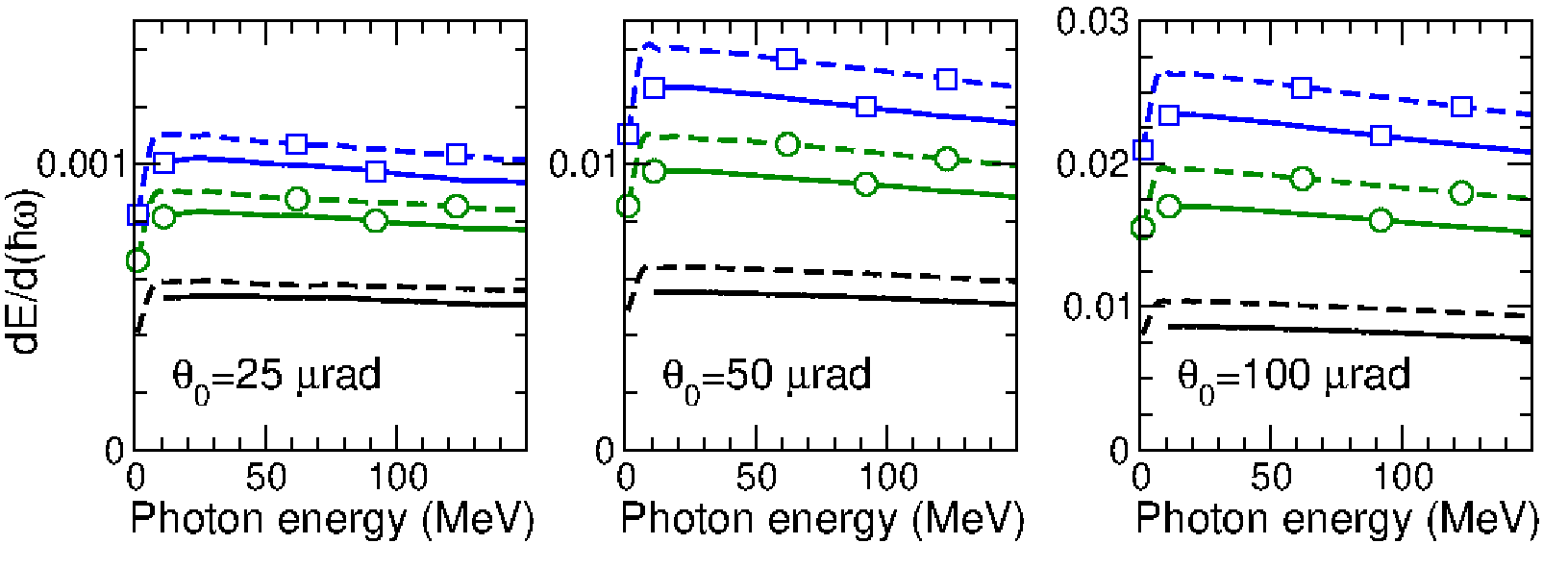}
\caption{
Spectral distribution of radiation (absolute values; note the factor
$10^3$ indicated in the axis label)
emitted by 10 GeV positrons in the amorphous
 diamond-like medium (solid lines) and amorphous
 silicon (dashed lines) environments.
The curves without symbols correspond to the crystal thickness
$L=1$ mm, the curves marked with open circles and squares -- to
$L=3$ and 6 mm, respectively.
Three graphs refer to different emission cones $\theta_0$ as
indicated.
}
\label{Figure_SM-01.fig}
\end{figure*}

Figure \ref{Figure_SM-01.fig} presents the
results of calculations for the spectral distributions
(absolute units multiplied by the factor $10^3$).
Three graphs refer to the emission cones
$\theta_0 =25$ $\mu$rad (left), 50 $\mu$rad (middle) and 100 $\mu$rad (right), which correspond approximately to
$1/2\gamma$, $1/\gamma$ and $2/\gamma$, respectively.
Solid curves show the dependencies calculated for the
amorphous diamond target,
dashed curves -- to the amorphous silicon target.
In each graph and for each type of the envitonment the results
shown correspond to
different values of the target's thickness: $L=1$ mm (curves without
symbols), 3 mm (open circles), 6 mm (open squares).

\begin{figure*} [h]
\centering
\includegraphics[width=16.4cm]{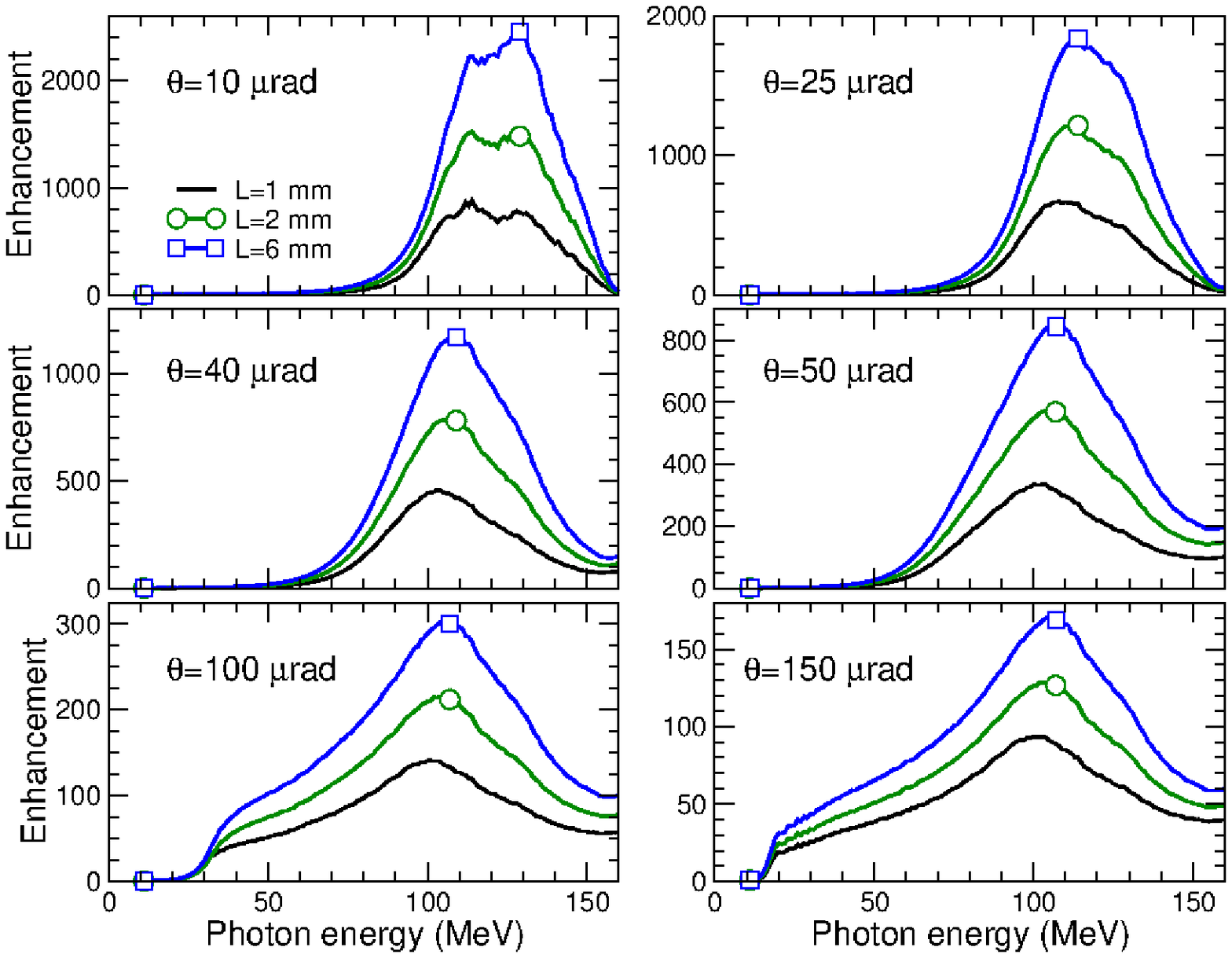}
\caption{
Enhancement factor of the channeling radiation in oriented
diamond (110) over the spectrum in amorphous diamond environment
calculated for 10 GeV positrons.
Each graph corresponds to the  emission cones $\theta_0$ indicated in its top left corner.
There curves (solid lines without symbols, with open circles and with
open squares) refer to different thicknesses $L$ of the targets
indicated in the common legend shown in the top left graph.
}
\label{Figure_SM-02.fig}
\end{figure*}

\begin{figure*} [h]
\centering
\includegraphics[width=16.4cm]{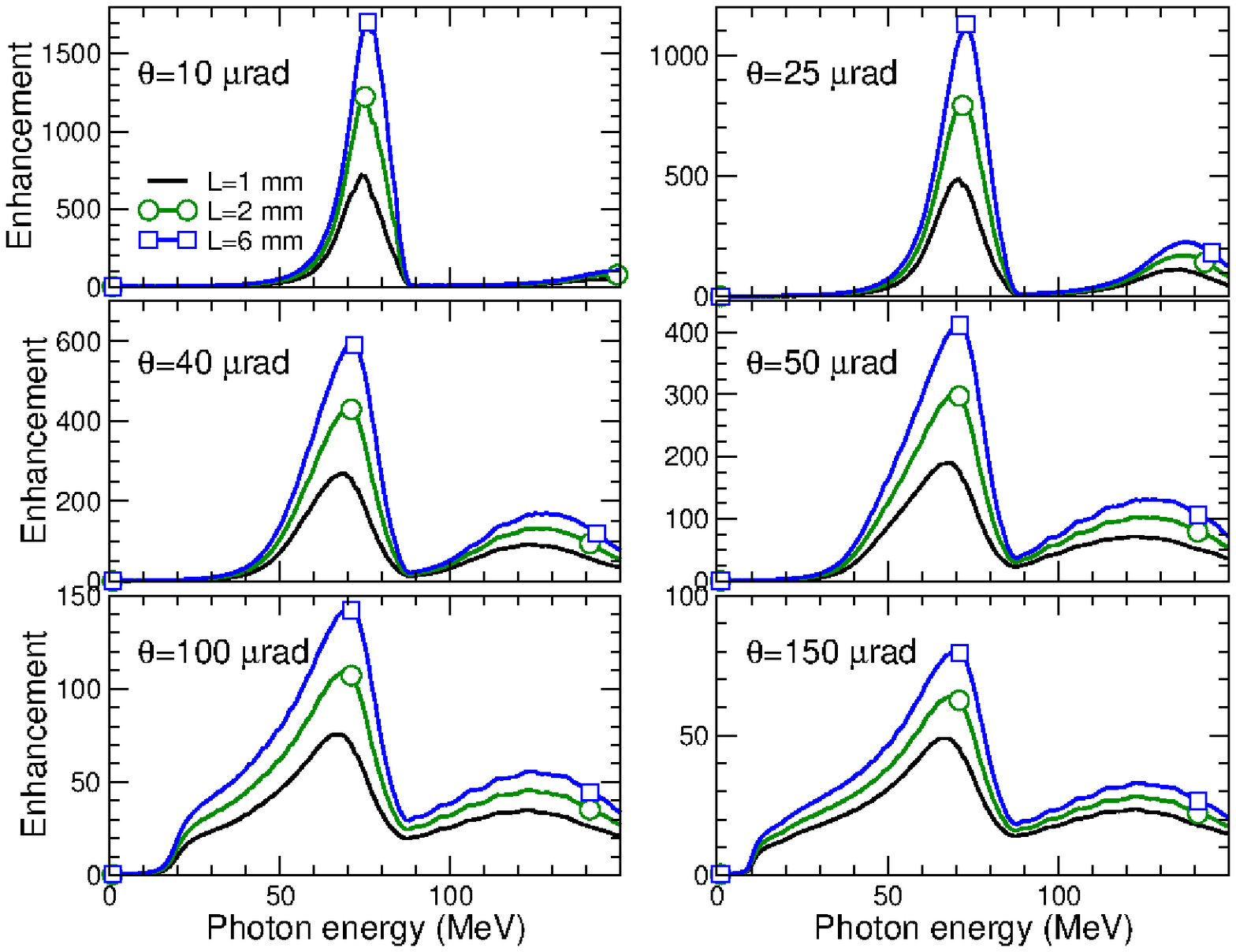}
\caption{
Same as Fig. (\ref{Figure_SM-02.fig}) but for the crystalline and amorphous silicon environments.
}
\label{Figure_SM-03.fig}
\end{figure*}

Enhancement factor, i.e. the ratio of the spectral intensity
$\d E/\d(\hbar\om)$ emitted in an oriented crystal to that
in the amorphous medium, is shown in
Figs. \ref{Figure_SM-02.fig}
(the diamond enviroment) and
\ref{Figure_SM-03.fig} (the silicon enviroment).
In each figure six graphs refer to the emission cones
indicated in the graphs' top left corner.
The dependences presented have been calculated for three
values of the targets' thickness $L$ in the direction of the incident
beam (see the common legends in the top left graphs in each figure).

\section{Harmonic approximation to the interplanar potential:
Estimates
\label{Estimates}}

The results presented and discussed in the main text have been obtained by means of
all-atom relativistic molecular dynamics approach implemented in the
\textsc{MBN Explorer}software
package \cite{MBNExplorer_2012,MBN_ChannelingPaper_2013,MBNExplorer_Book,
MBN_ChannelingPaper_2013}.
This approach goes beyond the continuous potential model \cite{Lindhard}.
Therefore, the numerical data presented below in this section are of an estimated nature.

Within the harmonic approximation, the continuous
interplanar potential
$U$ depends quadratically on the distance $\rho$ from the
channel's mid-plane:
\begin{eqnarray}
U(\rho)
= {m\gamma \Om_{\rm ch}^2 \rho^2 \over 2}\,,
\label{Estimates:eq.00}
\end{eqnarray}
where $\Om_{\rm ch}$ stands for the frequency of channeling oscillations
of an ultrarelativistic ($\gamma \gg1$) positron.
Within the harmonic approximation $\Om_{\rm ch}$ is independent on the
amplitude $a_{\rm ch}$ of the oscillations.

Notating the interplanar distance as $d$ and introducing the
depth of the interplanar potential $U_0=U(d/2)$
one estimates $\Om_{\rm ch}$
and spatial period  $\lambda_{\rm ch}$ of the
channeling oscillations as follows:
\begin{eqnarray}
\Om_{\rm ch} = \sqrt{8U_0 \over d^2 m\gamma}
=
{2c \over d} \sqrt{2U_0 \over \E},
\quad
\lambda_{\rm ch} \approx {2\pi c\over \Om_{\rm ch}}
= \pi d \sqrt{\E\over 2U_0}\,.
\label{Estimates:eq.01}
\end{eqnarray}

Directing the $z$ axis along the mid-plane and the
$y$ axis normally to the crystallographic plane,
and ignoring motion along the $x$ axis due
to the random collisions with the atoms,
one writes the trajectory of a channeling positron
as follows:
\begin{eqnarray}
y(z) = a_{\rm ch}\cos {2\pi z\over\lambda_{\rm ch}}
\label{Estimates:eq.02}
\end{eqnarray}
where $a_{\rm ch} \leq d/2$.
The undulator parameter
$K_{\rm ch} = 2\pi \gamma a_{\rm ch}/\lambda_{\rm ch}$
corresponding to this trajectory can be written as follows:
\begin{eqnarray}
K_{\rm ch}
= K_{\rm max}\left({2 a \over d}\right),
\quad
\mbox{with}\quad
K_{\rm max} = \gamma \sqrt{2U_0 \over \E}.
\label{Estimates:eq.03}
\end{eqnarray}
Hence, $K_{\rm ch}$ varies from $0$ at $a_{\rm ch}=0$
up to $K_{\rm max}$ for $a_{\rm ch}=d/2$.

The energy of the fundamental harmonic of the channeling radiation
emitted in \textit{the forward direction}
is estimated as follows
\begin{eqnarray}
\hbar \om\,\mbox
=
{2\gamma^2 \hbar\Om_{\rm ch} \over 1 + K_{\rm ch}^2/2}
\
\Longrightarrow
\
\left\{
\begin{array}{l}
\hbar \om_{\max}
=
2\gamma^2 \hbar\Om_{\rm ch}
\\
\displaystyle
\hbar \om_{\min}
=
{\hbar \om_{\max} \over 1 + K_{\max}^2/2}
\end{array}
\right.
\label{Estimates:eq.08}
\end{eqnarray}

\begin{table}
\caption{
Values of
$\theta_{\rm L}$ (in $\mu$rad),
$\lambda_{\rm ch}$ (in $\mu$m),
$K_{\max}$
and
$\hbar \om_{\min}$, $\hbar \om_{\max}$ (in MeV),
estimated for a
$\E=10$ and 0.5 GeV positron channeling in
C(110) ($d=1.26$ \AA{}, $U_0\approx 23$ eV) and
Si(110) ($d=1.92$ \AA{}, $U_0\approx 22$ eV).
}
\begin{tabular}{p{1.5cm}p{1.0cm}p{1.0cm}p{1.0cm}p{1.0cm}p{1.0cm}p{0.5cm}p{1.0cm}p{1.0cm}p{1.0cm}p{1.0cm}p{1.0cm}}
\hline
  &\multicolumn{5}{c}{$\E=10$ GeV} & &\multicolumn{5}{c}
  {$\E=0.53$ GeV}
  \\
  & $\theta_{\rm L}$&$\lambda_{\rm ch}$&$K_{\max}$&$\hbar\om_{\min}$ & $\hbar\om_{\max}$
  &
  & $\theta_{\rm L}$&$\lambda_{\rm ch}$&$K_{\max}$&$\hbar\om_{\min}$ & $\hbar\om_{\max}$
  \\
\hline
C(110) & 68 & 5.8 & 1.33& 84  & 159
&
& 295 & 1.3 & 0.31 & 1.7  & 1.9
\\
Si(110)& 66 & 9.1 & 1.31& 56  & 104
&
& 288 & 2.1 & 0.31 & 1.1  & 1.3
\\
\hline
 \end{tabular}
\label{Table.01}
\end{table}

 Table \ref{Table.01} summarizes the estimates obtained for
 a 10 GeV positron channeled in the (110) planar channels in
 diamond and silicon crystals.
 Also given are the values of Lindhard's critical angle
\begin{eqnarray}
\theta_{\rm L} = \sqrt{2U_0 \over \E}\,.
\label{Estimates:eq.04}
\end{eqnarray}

\section*{References}

\bibliography{references}

\end{document}